\documentclass[preprint,aps,showpacs,nofootinbib,preprintnumbers,amsmath,amssymb]{revtex4}
\usepackage{amssymb}
\usepackage{}
\usepackage{epsfig}
\usepackage{graphicx}
\usepackage{subfigure}
\usepackage{dcolumn}
\usepackage{bm}
\usepackage[usenames ,dvipsnames]{xcolor}
\usepackage{slashed}

\begin{document}

\title{Connecting Neutrino Masses and Dark Matter by High-dimensional Lepton Number Violation Operator}

\author{Chao-Qiang~Geng$^{1,2,3}$\footnote{geng@phys.nthu.edu.tw},
Da~Huang$^{2}$\footnote{dahuang@phys.nthu.edu.tw},
Lu-Hsing Tsai$^{2}$\footnote{lhtsai@phys.nthu.edu.tw}
and
Qing Wang$^{4,5}$\footnote{wangq@mail.tsinghua.edu.cn}
}
  \affiliation{$^{1}$Chongqing University of Posts $\&$ Telecommunications, Chongqing, 400065, China\\
  $^{2}$Department of Physics, National Tsing Hua University, Hsinchu, 300 Taiwan\\
  $^{3}$Physics Division, National Center for Theoretical Sciences, Hsinchu, 300 Taiwan\\
  $^{4}$Department of Physics, Tsinghua University, Beijing 100084, China\\
 $^5$Collaborative Innovation Center of Quantum Matter, Beijing 100084, China
}

\date{\today}

\begin{abstract}
We propose a new model with the Majorana neutrino masses generated at two-loop level, in which the lepton number violation (LNV) processes, such as neutrinoless double beta decays, are mainly induced by the dimension-7 LNV effective operator $\mathcal{O}_7=\bar l_R^c \gamma^\mu L_{L}(D_\mu \Phi) \Phi \Phi$. Note that it is necessary to impose an $Z_2$ symmetry in order that $\mathcal{O}_7$ dominates over the conventional dimension-5 Weinberg operator, which naturally results in a stable $Z_2$-odd neutral particle to be the cold dark matter candidate. More interestingly, due to the non-trivial dependence of the charged lepton masses, the model predicts the neutrino mass matrix to be in the form of the normal hierarchy. We also focus on a specific parameter region of great phenomenological interests, such as electroweak precision tests, dark matter direct searches along with its relic abundance, and lepton flavor violation processes.
\end{abstract}

\maketitle
\section{Introduction}
The presence of the tiny neutrino masses and mixings between different neutrino flavors have been established by many neutrino oscillation experiments~\cite{Anselmann:1992kc,Fukuda:1998mi,Ahmad:2002jz,Ahmad:2002ka,Ahn:2006zza,
Abe:2011sj,An:2012eh}, while more and more evidences are accumulated for the existence of dark matter (DM) over the last several decades, with the most precise measurement of its relic abundance by PLANCK~\cite{Ade:2013zuv,Adam:2015rua}. Both phenomena cannot be explained within the Standard Model (SM), thus providing us with two windows towards new physics beyond it. An interesting idea is to connect neutrinos and DM in a unified framework as many existing attempts in the literature (see {\it e.g.} Refs.~ \cite{Krauss:2002px,Ma:2006km,Aoki:2008av,Gustafsson:2012vj}). We would like to push this connection further in the present paper.

In order to understand the mass hierarchy problem in the neutrino sector, there are many models in the literature to naturally generate small Majorana neutrino masses such as the traditional Seesaw~\cite{TypeIseesaw1,TypeIseesaw2,TypeIseesaw3,TypeIseesaw4,TypeIseesaw5,typeIIseesaw1,typeIIseesaw2,typeIIseesaw3,
typeIIseesaw4,typeIIseesaw5,typeIIseesaw6,typeIIseesaw7,Foot:1988aq} and radiative mass generation mechanisms~\cite{Krauss:2002px,Ma:2006km,Aoki:2008av,Gustafsson:2012vj,Zee:1980ai,Zee:1985id,Babu:1988ki}. Most of them can be summarized as a specific realization of the conventional dimension-5 Weinberg operator. However, the generation of Majorana neutrino masses only requires the lepton number violation (LNV) by two units, and there exist many other equally legitimate LNV effective operators~\cite{Babu:2001ex,de Gouvea:2007xp,delAguila:2012nu,Angel:2012ug,Angel:2013hla,delAguila:2013zba, Aparici:2013xga,Cai:2014kra, Helo:2015fba}, which are composed of the SM fields but with higher scaling dimensions. From the effective field theory perspective, it is generically believed that these high-dimensional effective operators are subdominated by the Weinberg operator due to the suppression from the corresponding high powers of the large cutoff. In order for these operators to show up as the leading contributions, one usually needs to impose an additional symmetry on the model to break the usual scaling arguments. Furthermore, if this symmetry is kept unbroken, then the lightest symmetry-protected neutral particle would provide a perfect DM candidate. In this way, the symmetry connects Majorana neutrino masses and DM physics by the high-dimensional effective operators. Such a connection
has been already exemplified by some recent three-loop neutrino mass models~\cite{Gustafsson:2012vj,Geng:2014gua,Jin:2015cla,Geng:2015coa}, which realize the dimension-9 effective operator ${\cal O}_9 = \overline{l^c}_{R}l_{R} [(D_\mu \Phi) \Phi]^2$ with the DM embedded in the loop.

In this study, we focus on a specific dimension-7 LNV operator  $\mathcal{O}_7=\bar l_R^c \gamma^\mu L_{L}(D_\mu \Phi) \Phi \Phi$ ~\cite{delAguila:2012nu, delAguila:2013zba,Aparici:2013xga} and construct a UV complete model with an unbroken $Z_2$ symmetry to
accomplish  the above general arguments. In this model, Majorana neutrino masses arise radiatively at two-loop level, and neutrinoless double beta ($0\nu\beta\beta$) decays are dominated by a new ``long-range'' contribution \footnote{The definitions of ``short'' and ``long'' range contributions to $0\nu\beta\beta$ follow Refs.~\cite{Deppisch:2012nb,Bonnet:2012kh}.}, as the results of the existence of ${\cal O}_7$, while DM can also be embedded naturally as the lightest $Z_2$-odd neutral state.

This paper is organized as follows. In Sec.~II, we first describe the particle content and write down the relevant part of the Lagrangian for the model. We then calculate the two-loop neutrino masses and new contributions to the $0\nu\beta\beta$  decay rate in the model, with the emphasis on their relations to the high-dimensional effective operator ${\cal O}_7$. In Sec.~III, the constraints on the model are addressed from the electroweak (EW) precision tests, dark matter searches, and lepton flavor violating (LFV) processes. Finally, we give the conclusions in Sec.~IV.

\section{Generation of Neutrino Masses and $0\nu\beta\beta$ Decays}
\subsection{The Model}
\begin{figure}
\includegraphics[width=16cm]{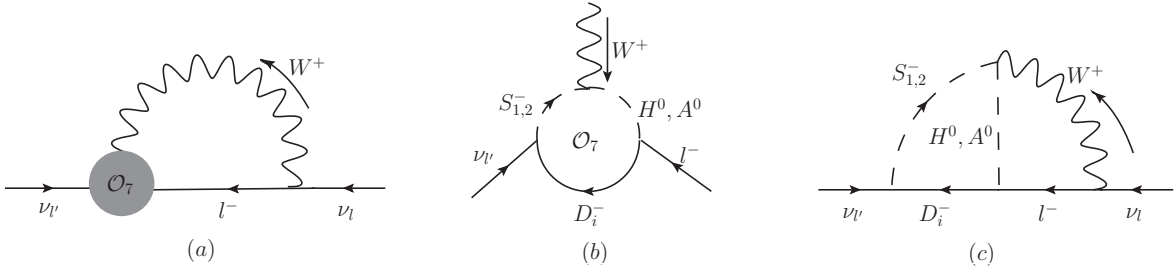}
\caption{Typical diagrams for (a) induced neutrino masses by the effective operator $\mathcal{O}_7$, (b) the one-loop realization of $\mathcal{O}_7$, and (c) the two-loop neutrino mass generation.}
\label{Fig_LNVdiagram}
\end{figure}

Fig.~\ref{Fig_LNVdiagram}a shows the neutrino mass generation induced by the one-loop diagram with $\mathcal{O}_7$.
In order to produce $\mathcal{O}_7$ at one-loop level, we introduce two scalars: $s:(1,2)$ and $\chi:(2,1/2)$, and three vector-like fermions $D_{(L,R)i}:(2,1/2)$ with $i=1,2$ and $3$ to the SM under ${\rm SU}(2)_L\times {\rm U}(1)_Y$.
A $Z_2$ symmetry is also imposed, in which only the new particles carry odd charges. The relevant new parts of the Lagrangian are given by
\begin{eqnarray}
-\Delta L&=&\mu_s^2(s^*s)+\mu_\chi^2(\chi^\dagger\chi)+V_4+\Big[\xi_{\,il} \overline {(D_{Li})^c}(i\sigma_2)L_{Ll} s+\zeta_{\,il} \overline {(D_{Li})} l_{Rl} \chi+ M_{Di} \bar D_{Li} D_{Ri}\nonumber\\
&&+\kappa s^*\chi^\dagger(i\sigma_2)\Phi+{\lambda_5\over2}(\chi^\dagger\Phi)^2+{\rm H.c.} \Big]\;,\label{Eq_Lagrangian}\\
V_{\rm 4}&=&\lambda_3(\Phi^\dagger\Phi)(\chi^\dagger\chi)
+\lambda_4(\Phi^\dagger\chi)(\chi^\dagger\Phi)+\lambda_{\Phi s}(\Phi^\dagger\Phi)(s^* s)
+\lambda_{\chi s}(\chi^\dagger\chi)(s^* s)\nonumber\\
&&+\lambda_\chi(\chi^\dagger\chi)^2
+\lambda_s(s^*s)^2\;,
\end{eqnarray}
where $\Phi=(\Phi^+,\,\Phi^0)^T$ is the SM Higgs doublet and $\sigma_2$ is the Pauli matrix for the SU(2)$_L$ gauge group. After the EW spontaneous symmetry breaking, $\Phi$ acquires a vacuum expectation value $v\equiv\sqrt{2}\langle \Phi^0\rangle\simeq 246\,{\rm GeV}$, while $\mu_\chi^2>0$ is necessary for preserving the $Z_2$ symmetry. Notice that the lepton number is explicitly broken only when $\kappa$, $\lambda_5$, and at least one of $\xi_{\,il}\zeta_{\,il'}$ are non-zero simultaneously. For convenience, we define $M_s^2=\mu_s^2+{1\over2}\lambda_{\Phi s}v^2$ and $M_\chi^2=\mu_\chi^2+{1\over2}\lambda_3v^2$. The trilinear coupling constant $\kappa$ in Eq.~(\ref{Eq_Lagrangian}) makes $s^\pm$ mix with $\chi^\pm$, which can be formulated as
\begin{eqnarray}
\left(\begin{array}{cc}
s^\pm\\
\chi^\pm\\
\end{array}\right)=
\left(\begin{array}{cc}
c_\theta&-s_\theta\\
s_\theta&c_\theta\\
\end{array}\right)
\left(\begin{array}{cc}
S_1^\pm\\
S_2^\pm\\
\end{array}\right)\;,\;
t_{2\theta}={\sqrt{2}\kappa v\over M_s^2-M_\chi^2}\;,
\end{eqnarray}
where $s_{x}\equiv \sin x$, $c_{x}\equiv \cos x$,  $t_{x}\equiv \tan x$, and
$S_1^\pm$ and $S_2^\pm$ represent the charged mass eigenstates with their masses, given by
\begin{eqnarray}
M_{S_1}^2&=&{c_\theta^2 \over c_{2\theta}}M_s^2-{s_\theta^2 \over c_{2\theta}}M_\chi^2\;,\;\nonumber\\
M_{S_2}^2&=&{c_\theta^2 \over c_{2\theta}}M_\chi^2-{s_\theta^2 \over c_{2\theta}}M_s^2\;,
\end{eqnarray}
respectively.
On the other hand, $\lambda_5$ contributes to the mass splitting between $H^0$ and $A^0$ in $\chi$, shown as
\begin{eqnarray}
M_{H}^2=M_\chi^2+{1\over2}(\lambda_4+\lambda_5) v^2\;,\;
M_{A}^2=M_\chi^2+{1\over2}(\lambda_4-\lambda_5) v^2\;.
\end{eqnarray}
As for the new fermions, we have the tree-level relation $M_{D_i^\pm}=M_{D_i^0}$ for each $D_i$ doublet.
The mass splittings between the charge and neutral components of the inert fermion doublets can only be induced by loop corrections with values around a few hundred MeV~\cite{Cirelli:2009uv}.

In this paper, we will characterize the model by using the physical quantities:
\begin{eqnarray}
M_H, M_A, M_{S_1}, M_{S_2}, M_{D_i},s_\theta, \lambda_L\,, \xi_{il},\;{\rm and} \;\zeta_{il},
\end{eqnarray}
where $\lambda_L\equiv {1\over2}(\lambda_3+\lambda_4+\lambda_5)$, and the other independent coupling constants from quarter terms:
\begin{eqnarray}
\lambda_{\Phi s}, \lambda_{\chi s}, \lambda_{\chi},\; {\rm and}\; \lambda_{s},
\end{eqnarray}
which are less relevant in our discussion.

\subsection{Two-Loop Majorana Neutrino Masses}
As seen in Fig.~\ref{Fig_LNVdiagram}b, the effective operator $\mathcal{O}_7$ can be induced by the one-loop diagram, whereas the Weinberg operator cannot\footnote{There is a similar realization of $\mathcal{O}_7$ in Ref.~\cite{Aparici:2013xga}, in which a triplet replaced the singlet $s$ of our model. A fundamental distinction of their paper from the present one is that ${\cal O}_7$ does not give the dominant contribution to Majorana neutrino masses in Ref.~\cite{Aparici:2013xga}. }.
Consequently, Majorana neutrino masses  appear through the two-loop diagram in Fig.~\ref{Fig_LNVdiagram}c\footnote{Similar topology with one $W^\pm$ exchange in a two-loop neutrino mass model can also be found in Refs.~\cite{Babu:2010vp,Babu:2011vb}, in which a different high-dimensional effective operator is realized without  DM.}. The resulting neutrino mass matrix $M_\nu$ defined in the Lagrangian $-{1\over 2}\overline{(\nu_L^c)}_l (M_\nu)_{ll'}(\nu_L)_{l'}+{\rm H.c.}$
 can be calculated as
\begin{eqnarray}
(M_\nu)_{ll'}=-{1\over (16\pi^2)^2}{G_F s_{2\theta}\over \sqrt{2}}(M_{H}^2-M_{A}^2)\sum_{i=1,2,3}(m_l\,\zeta_{il}\,\xi_{il'}+m_{l'}\,\zeta_{il'}\,\xi_{il})
[I_{i1}-I_{i2}]\;,\label{Eq_Mn}
\end{eqnarray}
where $m_l$ ($l=e,\,\mu,\,\tau$) are charged lepton masses, and $I_{ij}$ are defined by
\begin{eqnarray}
I_{ij}&\equiv& \int_0^1 dy_2\int_0^{1-y_2} dy_1\int_0^1 dx_3\int_0^{1-x_3} dx_2\int_0^{1-x_2-x_3} dx_1\nonumber\\
&&\Big\{\Big[{2(1-3x)\over x(1-x)}+{6y_1(2-x)\over (1-x)^2}\Big]\log( m_{ij}^2)+{-2 y_1(2-x)\over x(1-x)}{M_W^2\over m_{ij}^2}\Big\}\;,\\
m_{ij}^2&\equiv& y_1[x_1M_H^2+x_2M_A^2+x_3M_W^2]+y_2x(1-x)M_{S_j}^2\nonumber\\
&&+(1-y_1-y_2)x(1-x)M_{Di}^2\;,\nonumber\\
x&=&x_1+x_2\;.
\end{eqnarray}
Subsequently, one can diagonalize $M_\nu$ by
\begin{eqnarray}
{\rm diag}(m_1,m_2,m_3)=V^T M_\nu V,
\end{eqnarray}
where $m_{1,2,3}$ are three neutrino mass eigenvalues, which can have the normal ordering, $m_1<m_2\ll m_3$, or  inverted ordering,
 $m_3\ll m_1<m_2$, and $V$ is the Pontecorvo-Maki-Nakagawa-Sakata mixing matrix~\cite{Pontecorvo:1957cp,Maki:1962mu}. Without loss of generality, $V$ can be written as the standard parametrization by appropriate rephasing in $L_L$'s and $l_R$'s, given by~\cite{Agashe:2014kda}
\begin{eqnarray}
V=\left(\begin{array}{ccc}
c_{12}c_{13}&s_{12}c_{13}&s_{13}e^{-i\delta}\\
-s_{12}c_{23}-c_{12}s_{23}s_{13}e^{i\delta}&c_{12}c_{23}-s_{12}s_{23}s_{13}e^{i\delta}&s_{23}c_{13}\\
s_{12}s_{23}-c_{12}c_{23}s_{13}e^{i\delta}&-c_{12}s_{23}-s_{12}c_{23}s_{13}e^{i\delta}&c_{23}c_{13}\\
\end{array}\right)
\left(\begin{array}{ccc}
1&0&0\\
0&e^{i\alpha_{21}/2}&0\\
0&0&e^{i\alpha_{31}/2}\\
\end{array}
\right)\,,
\end{eqnarray}
where the mixing angles with $s_{ij}\equiv \sin\theta_{ij}$ and $c_{ij}\equiv \cos\theta_{ij}$ are defined within $[0,\,\pi/2]$, and Dirac phase $\delta$ and Majorana phases $\alpha_{21}$ and $\alpha_{31}$ are defined within $[0,2\pi]$.

From Eq.~(\ref{Eq_Mn}), we can get two important features for this mass generation mechanism. Firstly, the overall size of $M_\nu$ is proportional to the mass difference of the neutral scalars, $M_{H}^2-M_A^2$, and the combined factor of the charged states, $s_{2\theta}(I_{i1}-I_{i2})$, in which the former is generated by $\lambda_5$ and the latter corresponds to the size of $\kappa$. Turning off one of them will make all neutrinos massless. Secondly, the neutrino masses are positively correlated to the coupling matrix elements $\xi_{\,il}$ and $\zeta_{\,il}$, as well as the sizes of $m_l$. As the existence of the charged lepton mass hierarchy, $m_e\ll m_\mu < m_\tau$,
if  both matrices of $(\xi_{\,il})$ and $(\zeta_{\,il})$ are in uniform textures, the magnitude of $(M_\nu)_{ee}$ should be much smaller than those of other $M_\nu$ entries. We make a great advantage of this general expectation by taking the following limit
\begin{eqnarray}
(M_\nu)_{ee}&\simeq&0\;,\label{Mee}
\end{eqnarray}
which is shown in Refs.~\cite{Pascoli:2001by,Agashe:2014kda} to rule out the inverted ordering of neutrino mass spectrum at more than $2\sigma$ confident level. Thus, the normal ordering is predicted for the present model. Note that in the limit of Eq.~(\ref{Mee}), Ref.~\cite{Agashe:2014kda} even shows that the lightest neutrino mass $m_1$ can only be located within the range $0.001 {\rm eV}\lesssim  m_1 \lesssim 0.01{\rm eV}$.
Moreover, the smallness of $(M_\nu)_{ee}$
is also required by the $0\nu\beta\beta$ decay, which will be clear in the next subsection.

If we further focus on the CP conserving case, {\emph i.e.}, $\delta$, $\alpha_{21}$, $\alpha_{31} = 0$ or $\pi$, then $m_1$ will be constrained in the two narrow regimes, around $0.002$ and $0.007\,{\rm eV}$ with $\{\delta,\alpha_{21}, \alpha_{31}\}=\{0,\pi,0\}$ (Texture A) or $\{\pi,\pi,0\}$ (Texture B) and $\{0, \pi,\pi\}$ (Texture C) or $\{\pi, \pi,\pi\}$ (Texture D), respectively. Taking the central values of $\theta_{12}$, $\theta_{23}$, $\theta_{13}$, $\Delta m_{21}^2$, and $\Delta m_{32}^2$ from the global fitting for the neutrino oscillation data~\cite{Agashe:2014kda}, the corresponding mass matrices for Textures A, B, C, and D (${\rm T_A}$, ${\rm T_B}$, ${\rm T_C}$ and ${\rm T_D}$) are given by

\begin{eqnarray}
&&{\rm T_A}: M_{\nu}=
\left(\begin{array}{ccc}
0&0.12&0.92\\
0.12&1.9&2.7\\
0.92&2.7&2.4\\
\end{array}\right)(10^{-2})\,{\rm eV}\;,\;\label{Eq_MassTA}
\end{eqnarray}
\begin{eqnarray}
&&{\rm T_B}:M_{\nu}=
\left(\begin{array}{ccc}
0&-0.90&-0.24\\
-0.90&1.7&2.7\\
-0.24&2.7&2.6\\
\end{array}\right)(10^{-2})\,{\rm eV}
\,,\,
\label{Eq_MassTB}
\end{eqnarray}
\begin{eqnarray}
&&{\rm T_C}:M_{\nu}=
\left(\begin{array}{ccc}
0&-1.1&-0.055\\
-1.1&-2.3&-2.1\\
-0.055&-2.1&-3.1\\
\end{array}\right)(10^{-2})\,{\rm eV}\,,\,\label{Eq_MassTC}
\end{eqnarray}
\begin{eqnarray}
&&{\rm T_D}:
 M_{\nu}=
\left(\begin{array}{ccc}
0&-0.086&1.1\\
-0.086&-2.6&-2.2\\
1.1&-2.2&-2.9\\
\end{array}\right)(10^{-2})\,{\rm eV}
\,,
\label{Eq_MassTD}
\end{eqnarray}
respectively.

We now search for possible coupling matrix forms to realize the above four CP conserving neutrino textures. For simplicity, hereafter we will take $M_{D1}=M_{D2}=M_{D3}= M_{D}$, and also  set $\xi$ proportional to the identity matrix with the diagonal matrix element
to be $\xi_d$. Taking a symmetric form of $\zeta$,
the mass matrix element should be proportional to $\xi_d\zeta_{\,l'l}(m_l+m_{l'})$. We remark that by an appropriate phase absorption to the fermion fields, one can always have a positive $\xi_d$ without loss of generality. Comparing with Eqs.~(\ref{Eq_MassTA})-(\ref{Eq_MassTD}),
the forms of $\zeta_{\,ll'}$ in the four neutrino matrix textures can be obtained as
\begin{eqnarray}
&&{\rm T_A}: \zeta\propto
\left(\begin{array}{ccc}
{\rm \times}&0.12&0.052\\
0.12&0.89&0.14\\
0.052&0.14&0.068\\
\end{array}\right)\;,\;
{\rm T_B}: \zeta\propto
\left(\begin{array}{ccc}
{\rm \times}&-0.84&-0.013\\
-0.84&0.82&0.14\\
-0.013&0.14&0.072\\
\end{array}\right)
\,,\,
\nonumber\\
&&{\rm T_C}:\zeta\propto
\left(\begin{array}{ccc}
{\rm \times}&-1.&-0.0031\\
-1.&-1.1&-0.11\\
-0.0031&-0.11&-0.088\\
\end{array}\right)\,,\,
{\rm T_D}:
 \zeta\propto
\left(\begin{array}{ccc}
{\rm \times}&-0.081&0.062\\
-0.081&-1.2&-0.12\\
0.062&-0.12&-0.081\\
\end{array}\right)
\,,\nonumber\\
\label{Eq_textures}
\end{eqnarray}
where the cross means that the value of $\zeta_{\,ee}$ is still arbitrary at this stage, which will be constrained by $0\nu\beta\beta$ decays. The overall scale of $\zeta_{\,ll'}$ can be determined by Eq.~(\ref{Eq_Mn}) when all new particle masses, $s_\theta$ and $\xi_d$ are known.
We will also leave the discussion about the correlation between $\xi$ and $\zeta$ from the LFV constraints to Sec.~III.
\subsection{Neutrinoless Double Beta Decay}
In the previous section, we have built a two-loop neutrino mass model in which the LNV operator $\mathcal{O}_7$ provides the leading contribution. The next step is to study the LNV effect in this model induced by this high-dimensional operator. The most sensitive smoking gun for the LNV is the $0\nu\beta\beta$ decay, which will place the strong constraint on $\xi_d\zeta_{\,ee}$ and $(M_\nu)_{ee}$.
\begin{figure}
\includegraphics[width=16cm]{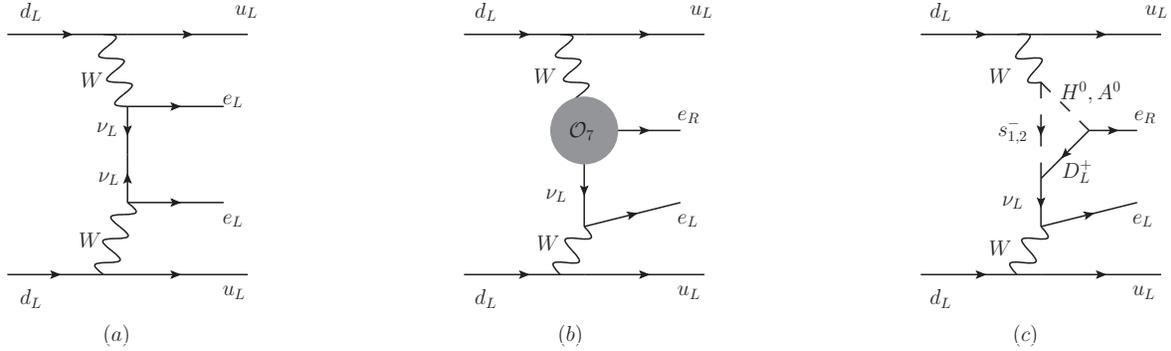}
\caption{Diagrams for $0\nu\beta\beta$ decays from (a) the traditional processes with the neutrino mass insertion on the propagator, (b) the contribution involving $\mathcal{O}_7$, and (c) the one-loop construction which realizes $\mathcal{O}_7$. For (b) and (c), the corresponding upside-down diagrams also need be considered.}
\label{Fig_0nbb}
\end{figure}

For the Majorana neutrino masses, there always exists the traditional long-range decay process by the exchange of neutrinos with a pair of left-handed electrons emitted as shown in Fig.~\ref{Fig_0nbb}a. Note that there is a chirality flipping on the internal neutrino propagator, which leads to the proportionality of the amplitude to the neutrino mass matrix element $(M_\nu)_{ee}$, in the sense that the detection for $0\nu\beta\beta$ processes could help to determine or constrain $|(M_\nu)_{ee}|$. However, it is known that in some types of neutrino models, Fig.~\ref{Fig_0nbb}a does not give the main part of this process, and one should have {\it a prior} consideration on the effects of other new diagrams. For example, a class of neutrino models~\cite{Chen:2006vn,Chen:2007dc,GT2015} that can be characterized by the dimension-9 operator $\mathcal{O}_9=\bar l_R^cl_R[(D_\mu\Phi)^T(i\sigma_2)\Phi][(D^\mu\Phi)^T(i\sigma_2)\Phi]$ is well studied in Refs.~\cite{Chen:2006vn,Chen:2007dc,delAguila:2011gr,Gustafsson:2014vpa,King:2014uha}, and the new contribution is much larger than that from Fig.~\ref{Fig_0nbb}a by orders of magnitude of $10^8$.
For our model or those with $\mathcal{O}_7$ as the main LNV source, the $0\nu\beta\beta$ decays are dominated by the diagram in Fig.~\ref{Fig_0nbb}b, which is not suppressed by the nearly-vanishing $(M_\nu)_{ee}$. We can write down a general formula for the half lifetime $T^{0\nu\beta\beta}_{1/2}$ of the $0\nu\beta\beta$  decay with the contributions from Figs.~\ref{Fig_0nbb}a and \ref{Fig_0nbb}b,  given by~\cite{Muto:1989cd}
\begin{eqnarray}
[T^{0\nu\beta\beta}_{1/2}]^{-1}&=&C_{mm}\Big({M_{ee}\over m_e}\Big)^2+C_{\eta\eta}\eta^2+C_{m\eta}\Big({M_{ee}\over m_e}\Big)\eta\;,\\
\eta&=&-{1\over 16\pi^2}{s_{2\theta}\xi\zeta_{\,ee}\over 4}(M_H^2-M_A^2)(I_{1}'-I_{2}')\;,\label{Eq_0nbb}
\end{eqnarray}
with
\begin{eqnarray}
I_{j}'&=&\int_0^1dx_3\int_0^{1-x3}dx_2\int_0^{1-x_2-x_3}dx_1{1\over x_1M_H^2+x_2M_A^2+x_3M_{S_j}^2+(1-x_1-x_2-x_3)M_{D}^2}\,,\nonumber
\end{eqnarray}
where $C_{mm}$, $C_{m\eta}$, and $C_{\eta\eta}$ include the phase space integrations and nuclear matrix elements defined in Ref.~\cite{Muto:1989cd}, and $\eta$ is the coupling of the interaction $(4G_F/\sqrt{2})(\bar u_L\gamma_\mu d_L)(\bar l_R\gamma^\mu \nu_L^c)$, which is originated from the one-loop generation for $\mathcal{O}_7$ (in Fig.~\ref{Fig_0nbb}c). By using the numerical results therein and also in Ref.~\cite{Doi:1985dx}, we find that the contribution proportional to $C_{\eta\eta}$ is much larger than those to $C_{mm}$ and $C_{m\eta}$.

As given in Eq.~(\ref{Eq_0nbb}) that $\eta$ is proportional to $\xi_d\zeta_{\,ee}$, the upper bound on $|\xi_d\zeta_{\,ee}|$ can be obtained by comparing with the current experimental sensitivities on $T^{0\nu\beta\beta}_{1/2}$~\cite{Agostini:2013mzu,KamLANDZen,Gando:2012zm,Argyriades:2008pr,Arnaboldi:2008ds,Arnold:2005rz,Barabash:2010bd} in Table~\ref{Tab_maxhy}, where we have used $M_H=70{\rm GeV }$, $M_A=95{\rm GeV }$, $M_{S_1}=310\,{\rm GeV}$, $M_{S_2}=90\,{\rm GeV}$, $M_D=200\,{\rm GeV}$, and $s_\theta=0.1$.
The strongest constraint is $|\xi_d\zeta_{\,ee}|<2.8\times10^{-4}$, given by the target nucleus Xe. Finally, the contribution from $\mathcal{O}_7$ (proportional to $C_{\eta\eta}$) is much larger than that from $(M_\nu)_{ee}$ (proportional to $C_{mm}$) by a factor of $\mathcal{O}(10^{-4})$,
 because the latter is greatly suppressed by the factor of $m_e/v$.
On the other hand, lifting up $|(M_\nu)_{ee}|$ to the average size of
$M_{\nu}\sim 10^{-2}~$eV  will result in the excess of the $0\nu\beta\beta$ decay rates that conflict with the observations.
 Table~\ref{Tab_maxhy} also shows the maximum value of $(M_\nu)_{ee}$ for each nucleus.

\begin{table}[ht]
\caption{Constraints on $|\xi_d\zeta_{\,ee}|$ from $0\nu\beta\beta$ for different nuclei as the targets. The corresponding limitation on $|M_\nu|_{ee}$ are also given.  }
\begin{tabular}{lccccl}
\hline
&$>T_{\rm exp}(10^{25}{\rm yr})$\;\;\;\;&$C_{\eta\eta}\,(10^{25}{\rm yr})^{-1}$&$|\xi_d\zeta_{\,ee}|_{\rm max}$&$|M_\nu|_{ee}(10^{-2}){\rm eV}$\\
\hline
GERDA-1($^{76}$Ge)~\cite{Agostini:2013mzu}&2.1&$4.4\times10^{-9}$&$3.6\times10^{-4}$&$<0.017$\\
KamLAND-Zen($^{136}$Xe)~\cite{KamLANDZen}&1.9&$8.3\times10^{-9}$&$2.8\times10^{-4}$&$<0.013$\\
NEMO-3($^{150}$Nd)~\cite{Argyriades:2008pr}&0.0018&$2.9\times10^{-7}$&$1.5\times10^{-3}$&$<0.072$\\
CUORICINO($^{130}$Te)~\cite{Arnaboldi:2008ds}&0.3&$2.3\times10^{-8}$&$4.2\times10^{-4}$&$<0.02$\\
NEMO-3($^{82}$Se)~\cite{Arnold:2005rz,Barabash:2010bd}&0.036&$1.5\times10^{-8}$&$1.5\times10^{-3}$&$<0.07$\\
NEMO-3($^{100}$Mo)~\cite{Barabash:2010bd}&0.11&$3.5\times10^{-8}$&$5.6\times10^{-4}$&$<0.027$\\
\hline
\end{tabular}
\label{Tab_maxhy}
\end{table}

Finally, we end this section by mentioning that {the role of
the $Z_2$ symmetry is} to make the dimension-7 operator ${\cal O}_7$ become the
dominant contributions to in the LNV processes and  Majorana neutrino masses.
Note that as the quantum numbers of the doublet ($\chi$) and the singlet ($s$) scalars are the same as those in the
Zee's model~\cite{Zee:1980ai},
the Majorana neutrino masses would be mainly generated
 by the corresponding one-loop diagrams related to
the conventional Weinberg operator if the $Z_2$ symmetry is absent.
However, with the $Z_2$ symmetry, ${\cal O}_7$ is singled out at 1-loop level, while other LNV effective operators, especially the dimension-5 Weinberg operator, are much suppressed since they would be only induced by higher-loop diagrams. In this way, the $Z_2$ symmetry breaks the conventional effective operator ordering based on the scaling dimensions. Other LNV effects, like $0\nu\beta\beta$ decays, would also change the leading modes accordingly.

\section{Phenomenological Constraints}

\subsection{Electroweak Precision Tests}
\begin{figure}
\includegraphics[width=16cm]{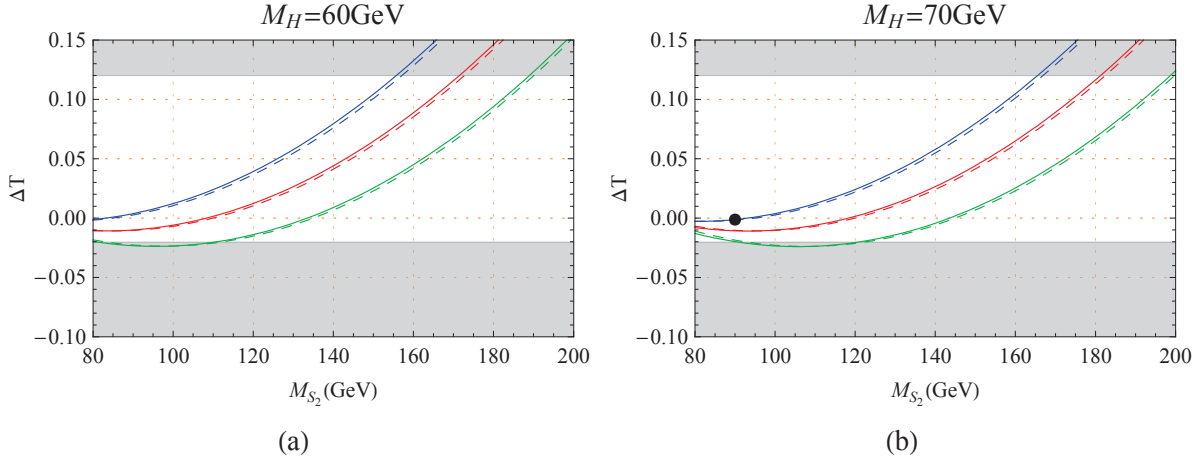}
\caption{Plots of $\Delta T$ versus $M_{S_2}$ with (a) $M_{H}=60\,{\rm GeV}$ and (b) $M_{H}=70\,{\rm GeV}$, where blue, red, and green colors represent $M_A-M_{H}=25$, $50$, and $75\,{\rm GeV}$, respectively, and solid (dashed) curve corresponds to $s_\theta=0.1\,(0)$, while the black dot is the benchmark point discussed in Sec.III.D.}
\label{Fig_Tpar}
\end{figure}

As discussed previously, in order to have the two-loop neutrino masses in our model, the non-zero coupling constants $\lambda_5$ and $\kappa$ are both required. The former splits the masses between $H^0$ and $A^0$, and the latter mixes the charged states $\chi^\pm$ and $s^\pm$ which carry different EW gauge quantum numbers. Both effects could change the values of the EW oblique $S$ and $T$ parameters. In particular, the $T$ parameter should yield a stronger constraint on this model.
The deviation of $T$ from the SM is given by~\cite{Gustafsson:2012vj}
\begin{eqnarray}
\Delta T&=&{1\over 4\pi s_W^2 M_W^2}\Big[{s_\theta^2\over4}(F_{S_1^\pm,H^0}+F_{S_1^\pm,A^0})+{c_\theta^2\over 4}(F_{S_2^\pm,H^0}+F_{S_2^\pm,A^0})-{1\over2}c_\theta^2 s_\theta^2F_{S_1^\pm,S_2^\pm}-{1\over4}F_{H^0,A^0}\Big]\;,\label{Eq_T}\nonumber\\
\end{eqnarray}
with the function $F$ defined by
\begin{eqnarray}
F_{x,y}={M_x^2+M_y^2\over 2}-{M_x^2 M_y^2\over M_x^2- M_y^2}\log\Big({M_x^2\over M_y^2}\Big)\;.
\end{eqnarray}
The value of $F_{x,y}$ becomes zero when $M_x\rightarrow M_y$, and
it increases with
 the mass splitting among the new scalars. Note that $\Delta T$ has little to do with $D_{i}$ since there is neither mixing between $D_i$ and the SM leptons nor tree-level mass splitting among $D_{i}$, while the deviation for the $S$ parameter can also be ignored~\cite{delAguila:2008pw}. The formulae of Eq.~(\ref{Eq_T}) is a general result for the models with the mixings between the inert doublet and singlet scalars.
The global fitting results constrain $\Delta T$ at $1.5-1.7\sigma$ deviation by $-0.02<\Delta T <0.12$~\cite{Agashe:2014kda}.
We show $\Delta T$ as a function of $M_{S_2}$ in Fig.~\ref{Fig_Tpar}, where we have used $M_{S_1}=310\,{\rm GeV}$ and $s_\theta=0.1(0)$ along with (a) $M_H=60$ and (b) $70\,{\rm GeV}$. It is obvious that the numerical result of our model with $s_\theta=0.1$ is approximately equal to a pure inert doublet model with $s_\theta=0$. In general, $\Delta T$ goes up with increasing $M_{S_2}$, and for a large value of $M_A-M_H$, the constraint on $M_{S_2}$ becomes stronger. The figure also shows that $M_A-M_H$ is limited to be less than $75\,{\rm GeV}$ for $M_H=60$ to $70\,{\rm GeV}$ and $90\lesssim M_{S_2}\lesssim110\,{\rm GeV}$. Finally, it should be noted that $M_{S_2}$ can not be too small, since there exists a lower bound on $M_{S_2}$ located within $70$ to $90\,{\rm GeV}$~\cite{Pierce:2007ut} from the LEP experiments.

\subsection{Dark Matter}
In this model, the lightest of the extra neutral particles: $H^0$, $A^0$, and $D^0_{1,2,3}$ could be a DM candidate, whose stability is guaranteed by the imposed $Z_2$ symmetry. In the following, we will focus on the case that DM is constituted solely by $H^0$ with a small charged scalar mixing $s_\theta$, in which our DM would be very similar to that in the well-studied inert doublet model~\cite{Barbieri:2006dq,Honorez:2010re,Arhrib:2013ela}. Furthermore, we concentrate on the low DM mass region with $50 ~{\rm GeV} \leqslant M_{H} \leqslant 80~{\rm GeV}$~\cite{Honorez:2010re,Arhrib:2013ela}, in which a large $H^0$-$A^0$ mass splitting can be allowed for the generation of the right two-loop neutrino masses. In addition, the mass of $S_2^\pm$ should be higher than 90~GeV in order to escape the LEP bounds~\cite{Pierce:2007ut}, so that the co-annihilation channels, such as $H^0$-$A^0$ and $H^0$-$S_2^\pm$, would be strongly suppressed and thus ignored.

\begin{figure}
\includegraphics[width=9cm]{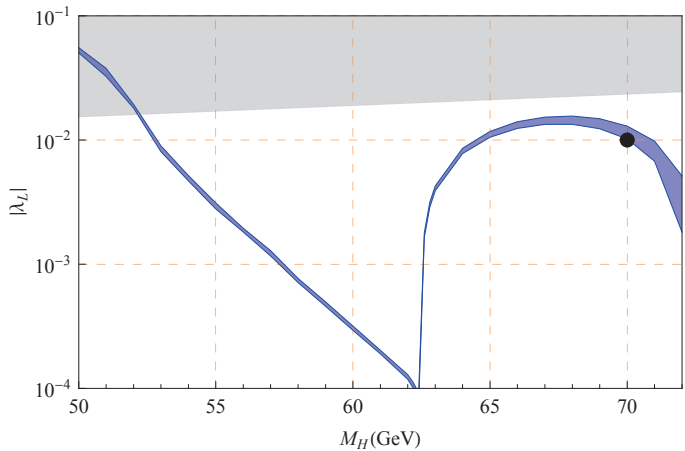}
\caption{The parameter space for the correction DM relic abundance in the $M_H-|\lambda_L|$ plane, with the blue band representing the region within $3\sigma$ deviation of the cold dark matter relic abundance. The gray shaded area is excluded by the LUX experiment, and the black dot represents the benchmark point. This plot is calculated with $M_{A}=95$~GeV and $M_{S_2}=90$~GeV.}
\label{Fig_DMRelic}
\end{figure}

We use the package 	
micrOMEGAs~\cite{Belanger:2014hqa} to accurately calculate the relic abundance $\Omega_H$ in the above parameter space, including all possible annihilations and co-annihilations. When $M_H$ approaches the half of the SM Higgs mass $M_h/2\simeq 62.5\,{\rm GeV}$~\cite{atlas:2012gk,cms:2012gu}, the Higgs resonance in the $s$-channel would become prominent, which is characterized by the coupling $\lambda_L$ controlling the trilinear vertex $(\lambda_L v)hH^0H^0$. However, in other regions, the DM annihilation cross section is dominated by the $WW^{(*)}$ mode. Therefore, the correct DM relic abundance is achieved mainly by the balance of the $WW^{(*)}$ and Higgs resonance channels. Fig.~\ref{Fig_DMRelic} shows the relevant parameter space in the $M_H$-$|\lambda_L|$ plane, which can give the observed DM abundance $0.112\lesssim\Omega_H h^2\lesssim0.128$ at $3\sigma$ level~\cite{Ade:2013zuv,Adam:2015rua,Agashe:2014kda}. Note that when  DM is heavier than 73~GeV, the $WW^{(*)}$ channel would give a too large annihilation cross section to accommodate the DM relic abundance~\cite{Honorez:2010re,Arhrib:2013ela}, which is omitted in Fig.~\ref{Fig_DMRelic}.

The DM $H^0$ in this low mass region could be constrained by the DM direct detection experiments. Since we need a relatively large Higgs-mediation annihilation channel to generate DM relic abundance, the Higgs exchange channel can also give rise to sizeable spin-independent signals, with the corresponding DM-nucleon cross section as follows~\cite{Honorez:2010re}:
\begin{eqnarray}
\sigma_{H^0 N}={m_r^2\over 4\pi}\Big({\lambda_L\over M_H M_h^2}\Big)^2f^2m_N^2\;.
\end{eqnarray}
Currently, the most stringent bound on the spin-independent DM-nucleon cross section is provided by the LUX experiment~\cite{Akerib:2013tjd}, with the minimum cross section of $7.6\times 10^{-46}~{\rm cm}^2$ for a DM mass of 33 GeV. It is shown in Fig.~\ref{Fig_DMRelic} that the LUX experiment has already probed some parameter space required by the DM relic abundance. Especially, the low DM mass region with $M_H \leqslant 52$~GeV is actually ruled out, as indicated by the shaded area in the plot. However, most parameter spaces are still allowed by LUX.

\subsection{Lepton Flavor Violation}
\begin{figure}
\includegraphics[width=16cm]{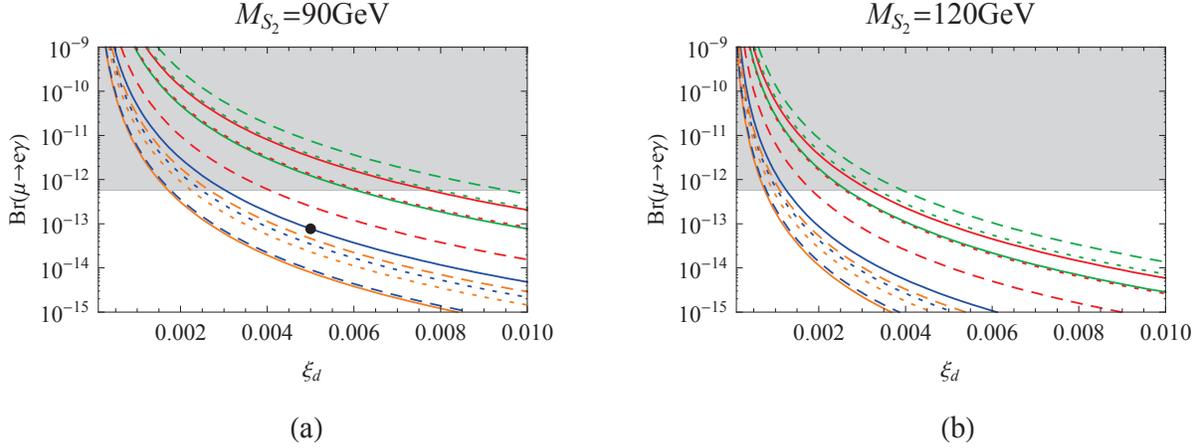}
\caption{${\rm Br}(\mu\rightarrow e\gamma)$ versus $\xi$, with (a) $M_{S_2}=90\,{\rm GeV}$ and (b) $M_{S_2}=120\,{\rm GeV}$, where the blue, red, green, and yellow curves correspond to the neutrino textures $T_A$, $T_B$, $T_C$, and $T_D$, respectively, while the black dot is the benchmark point.}
\label{Fig_m2egamma}
\end{figure}

The current experimental constraints on LFV processes, such as the radiative decays $l\rightarrow l'\gamma$~\cite{Adam:2013mnn,Aubert:2009ag}, $\mu-e$ conversions~\cite{Bertl:2006up,Badertscher:1980bt,Dohmen:1993mp,Honecker:1996zf}, and three-lepton decays $l\rightarrow l_1l_2\bar l_3$~\cite{Bellgardt:1987du,Hayasaka:2010np},  are all dominated by one-loop diagrams with $Z_2$ odd particles inside. We take $\mu\rightarrow e\gamma$ as an illustration because the current experimental upper bound ${\rm Br}(\mu^+\rightarrow e^+\gamma)<5.7\times 10^{-13}$~\cite{Adam:2013mnn} usually constrains a model in the most stringent way. In our model, we have
\begin{eqnarray}
{\rm Br}(\mu^\mp\rightarrow e^\mp\gamma)&=& {\Gamma(\mu^\mp\rightarrow e^\mp\gamma)\over \Gamma(\mu^\mp\rightarrow e^\mp\nu\bar\nu)}\nonumber\\
&=&{3\alpha_e\over 64\pi G_F^2}
\Big|\sum_{l} \zeta_{\,l\mu}\zeta_{\,le}\Big| ^2
\Big(s_\theta^2K_{S_1}+c_\theta^2
K_{S_2}+{1\over2}K_{H}'+{1\over2}K_{A}'\Big)^2\;,\label{Eq_m2eg}
\end{eqnarray}
where the loop integrals $K_{x}$ and $K_{x}'$ are defined as
\begin{eqnarray}
K_{x}&=&{2z^3+3z^2-6z+1-6z^2\log z\over 6(1-z)^4M_x^2}\,,\,
K_{x}'=-{z^3-6z^2+3z+2+6z\log z\over 6(1-z)^4M_x^2}\,,
\end{eqnarray}
with $z=M_{D}^2/ M_x^2$.
As is expected, the decay branching ratio is proportional to the coupling constant combination $\Big|\sum_{l} \zeta_{\,l\mu}\zeta_{\,l e}\Big| ^2$. On the other hand, when the values of $\xi_d\zeta_{\,ll'}$ are fixed by the neutrino masses, the only degree of freedom left is the size of $\xi_d$.
When $\xi_d$ is large, $\zeta_{\,ll'}$ should be suppressed, along with all the relevant LFV processes. This feature can be displayed in Fig.~\ref{Fig_m2egamma}, where $\zeta_{\,ll'}$ are expressed in the forms given in Eq.~(\ref{Eq_textures}), with the unknown $\zeta_{\,ee}$ satisfying the relation $|\xi_d\zeta_{\,ee}|\lesssim10^{-4}$, as well as $s_\theta=0.1$, $M_H=70$, $M_A=95$, $M_{S_1}=310$, $M_{S_2}=90$ and $120$ and $M_D=200\,{\rm GeV}$. We find that the texture $T_{C(D)}$ yields the most stringent (weakest) constraint on $\xi_d$, such that $\xi_d \gtrsim 0.01\,(0.002)$ is required for $M_{S_2}=90\,{\rm GeV}$. From another angle, if we set $\xi_d=0.005$ with $T_A$, we can predict that ${\rm Br}(\mu\rightarrow e\gamma)=10^{-13}\,(3\times 10^{-15})$ for $M_{S_2}=90\,(120)\,{\rm GeV}$, which might be measured by the next-generation experiments in the future.

\subsection{Numerical Results}
Based on the above constraints from the LFV processes, EW precision tests, direct searches of DM with the required relic abundance, we find a benchmark point from the allowed parameter space, given by:

\begin{eqnarray}
M_H=70\,{\rm GeV}\;,\;M_A=95\,{\rm GeV}\;,\;M_{S_1}=310\,{\rm GeV}\;,\; M_{S_2}=90\,{\rm GeV}\;,\nonumber\\
M_{D_1}=M_{D_2}=M_{D_3}=M_D=200{\rm GeV}\;,s_\theta=0.1\;,\;\lambda_L=0.01\,,
\end{eqnarray}
and
\begin{eqnarray}
\xi=\left(\begin{array}{ccc}
0.005&0.&0.\\
0.&0.005&0.\\
0.&0.&0.005\\
\end{array}\right)\,,\;\;
\zeta=
\left(\begin{array}{ccc}
0.02&0.005&0.0022\\
0.005&0.038&0.0061\\
0.0022&0.0061&0.0029\\
\end{array}\right)\,.
\end{eqnarray}
It is clear that the matrix $\zeta$ corresponds to the neutrino texture $T_A$. We also plot this benchmark point by a black dot in Figs.~\ref{Fig_m2egamma}a, \ref{Fig_Tpar}b, and \ref{Fig_DMRelic}, where the experimental results from LFV processes, oblique parameters, and DM searches are well satisfied, respectively. Finally, from the benchmark point, one can obtain $\lambda_3=0.155$, $\lambda_4=-0.067$, $\lambda_5=-0.068$, $\kappa=50.3\,{\rm GeV}$, $\mu_\chi=65.5\,{\rm GeV}$, and $M_S=309\,{\rm GeV}$, which are all small enough to ensure the self-consistence of the perturbation theory.

\section{Conclusions}
We have tried to make the connection between neutrino physics and dark matter searches. In particular, we have emphasized that every effective operator, which violates the lepton number by two units, can give an equally good mechanism to generate Majorana neutrino masses. The problem lies in the fact that the new high-dimensional operators might be buried by the overwhelming effects from the conventional Weinberg operator which possess the smallest scaling dimension. One way to break this effective field theory ordering is to impose some symmetry which would protect the lightest neutral symmetry-protected states to be the DM particle.

We have explicitly realized this connection by constructing a UV complete model with the $Z_2$ symmetry to
 generate the dimension-7 operator ${\cal O}_7 = \bar l_R^c \gamma^\mu L_{L}(D_\mu \Phi) \Phi \Phi$. We have shown that the Majorana neutrino mass matrix structure and the leading $0\nu\beta\beta$ decay contribution are closely related to ${\cal O}_7$. Especially, the neutrino mass matrix is predicted to be of the normal ordering due to the hierarchy in the charged lepton masses, and the $0\nu\beta\beta$ decay rate can be large enough  to be tested in the next-generation experiments. If we impose an additional CP symmetry in the lepton sector, we can even determine the form of the neutrino mass matrix completely. We have also focused on a specific parameter region with a small mixing between charged scalars, and considered the constraints from the electroweak precision tests, dark matter searches, and LFV processes.

\begin{acknowledgments}
This work was supported by National Center for Theoretical Sciences, National Science
Council (Grant No. NSC-101-2112-M-007-006-MY3) and National Tsing Hua University (Grant No. 104N2724E1), 
the National Science Foundation of China (NSFC Grant No. 11475092),  the Tsinghua University Initiative Scientific Research Program (Grant No. 20121088494).
\end{acknowledgments}

\end{document}